\documentclass[11pt]{article}
\newcommand{\Comment}[1]{{}}
\usepackage{amsfonts,amsthm,amsmath,amssymb,slashed}
\usepackage[textwidth = 430 pt, textheight = 630 pt]{geometry}
\usepackage{color}
\usepackage[section]{placeins}
\usepackage{comment}

\Comment{\usepackage{color}
\definecolor{MyDarkBlue}{rgb}{0.15,0.15,0.45}
\usepackage[linktocpage=true]{hyperref}
\hypersetup{
colorlinks=true,
citecolor=MyDarkBlue,
linkcolor=MyDarkBlue,
urlcolor=MyDarkBlue,
pdfauthor={Horatiu Nastase and Jacob Sonnenschein},
pdftitle={The title},
pdfsubject={hep-th}
}

\usepackage[numbers,sort&compress]{natbib}
\usepackage{hypernat}}
\usepackage{graphicx}
\usepackage{cite}
\usepackage{url}

\newcommand\ignore[1]{}
\def\one{{\,\hbox{1\kern-.8mm l}}}

\def\Tr{{\rm Tr\, }}

\def\b{\beta}

\def\d{\partial}

\def\Tr{\mathop{\rm Tr}\nolimits}

\newcommand{\Cset}{{\,\,{{{^{_{\pmb{\mid}}}}\kern-.45em{\mathrm C}}}}}

\newcommand{\be}{\begin{equation}}
\newcommand{\bea}{\begin{eqnarray}}

\newcommand{\ee}{\end{equation}}
\newcommand{\eea}{\end{eqnarray}}

\begin{document}

\renewcommand{\thefootnote}{\fnsymbol{footnote}}

\makeatletter
\@addtoreset{equation}{section}
\makeatother
\renewcommand{\theequation}{\thesection.\arabic{equation}}

\rightline{}
\rightline{}
%   \vspace{1.8truecm}

%\begin{flushright}
% preprint nrs.
%\end{flushright}

%\vspace{10pt}

%\begin{document}
\begin{center}
{\LARGE \bf{\sc Holographic Krylov complexity in ${\cal N}=4$ SYM}} 
\end{center} 
 \vspace{1truecm}
\thispagestyle{empty} \centerline{
{\large \bf {\sc Ali Fatemiabhari${}^{a},$}}\footnote{E-mail address: \Comment{\href{alifatemiabhari@gmail.com}}
{\tt alifatemiabhari@gmail.com}}
{\large \bf {\sc Horatiu Nastase${}^{b}$}}\footnote{E-mail address: \Comment{\href{mailto:horatiu.nastase@unesp.br}}
{\tt horatiu.nastase@unesp.br}}
%{\large \bf {\sc Carlos Nunez${}^{c},$}}\footnote{E-mail address: \Comment{\href{mailto:c.nunez@swansea.ac.uk}}
%{\tt c.nunez@swansea.ac.uk}} }
%\centerline{
{\bf{\sc and}}
{\large \bf {\sc Dibakar Roychowdhury${}^{c}$}}\footnote{E-mail address: \Comment{\href{mailto:dibakarphys@gmail.com}}{\tt dibakar.roychowdhury@ph.iitr.ac.in}}
                                                       }

\vspace{.5cm}

%\vspace{.3cm}

\centerline{{\it ${}^a$Institute for Theoretical and Mathematical Physics, }}
\centerline{{\it Lomonosov Moscow State University, 119991 Moscow, Russia}} 
%\centerline{{\it University of Cape Town, Cape Town, South Africa}}
\vspace{.3cm}
\centerline{{\it ${}^b$Instituto de F\'{i}sica Te\'{o}rica, UNESP-Universidade Estadual Paulista}} 
\centerline{{\it R. Dr. Bento T. Ferraz 271, Bl. II, Sao Paulo 01140-070, SP, Brazil}}
\vspace{.3cm}
%\centerline{{\it ${}^c$Department of Physics, Swansea University,}} 
%\centerline{{\it Swansea SA2 8PP, United Kingdom}}
%\vspace{.3cm}
\centerline{{\it ${}^c$Department of Physics, Indian Institute of Technology Roorkee,}}
\centerline{{\it Roorkee 247667, Uttarakhand, India}} 
 
\vspace{1truecm}

%%%%%%%%%%%%%%%%%
\thispagestyle{empty}

\centerline{\sc Abstract}

\vspace{.4truecm}

\begin{center}
\begin{minipage}[c]{380pt}
{\noindent 
We propose and calculate a holographic Krylov complexity in ${\cal N}=4$ SYM via the proper momentum 
for motion in $AdS_5$ sliced by $AdS_3$. The motion in an $AdS_3$ subgroup corresponds to the Krylov 
complexity of the $Sl(2)$ subsector. The general motion corresponds to the Krylov complexity of the 
${\cal N}=4$ SYM.
}
\end{minipage}
\end{center}

\vspace{.5cm}

\setcounter{page}{0}
\setcounter{tocdepth}{2}

\newpage

\tableofcontents
\renewcommand{\thefootnote}{\arabic{footnote}}
\setcounter{footnote}{0}

\linespread{1.1}
\parskip 4pt

%{}~
%{}~

%---------------------------------------------------------

%%%%%%%%%%%%%%%%%%%%%%%%%%%%%%%%%%%%%%%%%%%%%%%%%%%%%%%%%%%%%%%%%%%%%%%%%%%%%%%%%%%%%%%%

\section{Introduction}

The notion of complexity in quantum mechanical systems is a recent and one that has received a lot of 
interest. The geometrical description of Nielsen \cite{Nielsen:2005mkt} was followed by extensions via 
holography to quantum field theories, for the subsequent conjectures of ``complexity = volume'' 
\cite{Stanford:2014jda}, ``complexity = action'' \cite{Brown:2015bva} or, more recently, a one-dimensional 
set of conjectures dubbed ``complexity = anything'' \cite{Belin:2021bga}, and it has then led to understanding 
in quantum field theory of the notion of circuit complexity \cite{Jefferson:2017sdb}. But these notions
of complexity are not algorithmic. A more recent notion of complexity that {\em is} algorithmic, at least in 
quantum mechanics (0+1 dimensional systems) is Krylov complexity, initially defined as a measure of 
quantum chaos \cite{Parker:2018yvk} (for reviews of Krylov complexity, see 
\cite{Nandy:2024evd,Rabinovici:2025otw}, and 
for a review of all notions of quantum complexity, see \cite{Baiguera:2025dkc}), and then defined in a 
more general way, equivalently for states and operators, as ``spread complexity'' in 
\cite{Balasubramanian:2022tpr}.

The holographic description of Krylov complexity has been more difficult to define. 
In one line of research, for the quantum mechanical double-scaled SYK model (DSSYK), the holographic 
complexity was defined from the wormhole length in the dual JT gravity 
\cite{Rabinovici:2023yex,Ambrosini:2024sre} (see also the review \cite{Rabinovici:2025otw}). In another 
line of research, the time derivative of the Krylov complexity $\dot C(t)$ was shown to be related to the {\em 
proper momentum} $P_{\bar \rho}(t)$ 
for a motion in an $AdS_3$ gravity dual (so, for a $CFT_2$), a relation made precise 
in \cite{Caputa:2024sux}, as 
\be
\dot C(t)=-\frac{P_{\bar \rho}}{\epsilon}\;,\label{compmom}
\ee
based on various earlier, but less precise relations. Such a less precise relation was proven from the 
``complexity = volume'' conjecture for the Nielsen complexity in \cite{Barbon:2020uux} and previous work 
by the same authors. Note also the extension of Krylov complexity to supersymmetric theories, and to 
semiclassical strings in a gravity dual defined in \cite{Das:2024tnw}.

In this paper we seek to extend the range of the proposal in (\ref{compmom}) to the original case of 
holographic duality,
${\cal N}=4$ SYM vs. $AdS_5\times S^5$. We will further extend this in two companion papers to appear
shortly, one where 
we will extend this to the case of 
non-conformal confining theories, with only asymptotically AdS, but the space ending in the IR, 
based on the Anabalon-Ross solution \cite{Anabalon:2021tua} (related also to the generalization to the 
non-singular version of the Coulomb branch of ${\cal N}=4$ SYM in \cite{Anabalon:2024che}), itself 
generalizing \cite{Horowitz:1998ha}, and another where we will extend this to the case of 
$AdS_3$ and $AdS_2$ solutions, but with an extra direction corresponding to a quiver in the field 
theory \cite{Lozano:2019emq,Lozano:2019jza,Lozano:2019zvg,Lozano:2019ywa,Lozano:2020bxo}
and \cite{Lozano:2020txg,Lozano:2020sae,Lozano:2021rmk}, respectively.

In this ${\cal N}=4$ SYM case, we will first note that there is an $AdS_3$ subspace construction, dual 
to the $Sl(2)$ subset of ${\cal N}=4$ SYM, where previous constructions apply, and then generalize to 
the case of a slicing of $AdS_5$ by $AdS_3$, and propose a holographic construction that predicts 
a field theory Krylov complexity. 

The paper is organized as follows. In section 2 we review Krylov complexity and better define our 
proposal for its holographic form. In section 3 we consider the $AdS_3$ subspace of $AdS_5$ and its 
dual $Sl(2)$ subsector of ${\cal N}=4$ SYM. In section 4 we consider the general case, of slicing $AdS_5$
by $AdS_3$ slices, calculating the motion starting at zero velocity in the UV, and the resulting proper 
momentum. In section 5 we conclude.

\section{Krylov complexity and holographic proposal}

For a quantum mechanical system, the Krylov complexity is defined as the average number
(or discrete ``length'', the equivalent of the number of gates in a circuit for a quantum system) $n$ 
in a basis of the Hilbert space, the Krylov basis $|n\rangle$, to go from a reference state $|0\rangle$
to a target space $|\psi(t)\rangle$. Using the Lanczos algorithm, from $|0\rangle$ one defines the 
Krylov basis $|n\rangle$, in which the (Hermitian!) Hamiltonian has the most general tri-diagonal form, 
\be
\hat H|n\rangle=a_n|n\rangle+b_n|n-1\rangle +b_{n+1}|n+1\rangle\;,\label{HKrylov}
\ee
or the time-dependent target space obeys the (Schr\"{o}dinger) equation
\be
i\d_t\psi_n(t)=a_n\psi_n(t)+b_n\psi_{n-1}(t)+b_{n+1}\psi_{n+1}(t)\;,
\ee
where $\psi_n(t)\equiv \langle n|\psi(t)\rangle$. The Krylov complexity is then 
\be
C(t)\equiv \sum_n n|\psi_n(t)|^2=\sum_n n p_n(t)=\langle n\rangle(t).
\ee

In \cite{Caputa:2024sux}, the calculation of the Krylov (or spread) complexity of states excited by primary 
operators ${\cal O}$ of dimension $\Delta$ and energy $E_0=\Delta/\epsilon$ (with regulator $e^{-\epsilon H}$)
in a generic $CFT_2$ 
was done, in the cases of the CFT on the cylinder $\mathbb{R}_t\times S^1_L$, so for a CFT on a finite 
size $L$, and the CFT on the plane $\mathbb{R}_t\times \mathbb{R}_x$, so for a CFT on an infinite line, 
obtaining
\bea
\dot C_L(t)&=&\frac{E_0}{\epsilon}\frac{L}{2\pi}\sin \frac{2\pi t}{L}\Rightarrow \label{KrylovL}\\
\dot C_\infty(t)&=&\frac{E_0}{\epsilon} t\;\;({\rm for \;\;L\rightarrow\infty})\;,\label{Krylovinf}
\eea
respectively, and also at finite temperature $T=1/\b$, obtaining 
\be
\dot C_T(t)=\frac{E_0}{\epsilon}\frac{\b}{2\pi}\sinh \frac{2\pi t}{\b}\;,
\ee
consistent in the $T\rightarrow 0$ limit with the $C_\infty (t)$ case. These calculations were matched by 
the calculations in global $AdS_3$ (with $\mathbb{R}_t\times S^1$ boundary) and Poincar\'{e} 
$AdS_3$ (with $\mathbb{R}_t\times \mathbb{R}_x$ boundary), 
\bea
ds_{\rm 3,gl.}^2&=& d\rho^2 +\frac{4\pi^2}{L^2}(-\cosh^2 \rho \,dt^2+\sinh^2\rho\, d\phi^2)\cr
ds_{\rm 3,P.}^2&=& d\rho^2+e^{2\rho}(-dt^2+dx^2)\;,
\eea
respectively, where we have used the coordinates $\rho$ such that $ds^2=d\rho^2$ when the boundary 
coordinates are constant. The canonical momentum $P_\rho$ in these coordinates was referred to as the 
{\em proper momentum}. By starting in the UV, i.e., at very large $\rho$, with $z=e^{-\rho}=\epsilon$, 
and calculating the geodesic of a massive particle starting at zero velocity, one finds
in both cases that 
\be
\dot C(t)=-\frac{P_\rho(t)}{\epsilon}\;,
\ee
as advertised in (\ref{compmom}), with $\bar\rho=\rho$ here.
Similarly, at finite temperature $T$, for the $AdS_3$ black hole of temperature $T=1/\b$, 
\be
ds_{\rm 3,BH}^2=d\rho^2+\frac{4\pi^2}{\b^2}(-\sinh^2\rho \, dt^2+\cosh^2\rho\,dx^2)\;,
\ee
which has a horizon at $\rho=0$, that gives a periodicity of Euclidean time $\Delta t_E=\b=1/T$,
one finds the same relation to be satisfied.

The isometry of  $AdS_3$ is $SO(2,2)\simeq SO(2,1)\times SO(2,1)\simeq 
Sl(2;\mathbb{R})\times Sl(2;\mathbb{R})$, with the Lorentz group being $SO(2,1)\simeq Sl(2;\mathbb{R})$, 
and then $AdS_3$, which is the coset manifold $SO(2,2)/SO(2,1)\simeq Sl(2;\mathbb{R})$, is isometric 
to the noncompact group manifold $Sl(2;\mathbb{R})$, where the two $Sl(2;\mathbb{R})$ symmetries
are left and right actions on the group elements. 

This $Sl(2;\mathbb{R})$ symmetry structure is at the core of the solvability of Krylov complexity in 
this case, dual to $CFT_2$, and the sinusoidal form in the finite size $L$ case is due to the finite size of the
Hilbert space, if discretized by $\epsilon$ down to a spin chain form.

We are proposing that even in a non-$AdS_3$ situation, the formula (\ref{compmom}) continues to 
hold, if we identify correctly the {\em proper distance} coordinate $\bar \rho$, such that 
$ds^2=d\bar\rho^2$ {\em on the motion of the massive particle  falling from large $\bar\rho$}, and then 
$P_{\bar\rho}$ is the canonical momentum of this coordinate.

Specifically, then, we write on the timelike geodesic parametrized as $(r(t),\rho(t))$, that
\be
ds^2=A(t)dr(t)^2+B(t)d\rho(t)^2\equiv d\bar\rho(t)^2\;,
\ee
so that
\be
\frac{d\rho^2}{dr^2}=\frac{\dot\rho^2}{\dot r^2}\;,
\ee
and the proper momentum is 
\be
P_{\bar\rho}=P_r\frac{dr}{d\bar \rho}+P_\rho\frac{d\rho}{d\bar \rho}.
\ee

\section{$AdS_3$ subspace vs. $Sl(2)$ sector}

The first thing to understand is that there is an $AdS_3$ subspace inside $AdS_5$, and it is dual to 
the so-called $Sl(2)$ sector, and thus the complexity within this sector is given by the same 
calculation as the generic $CFT_2$ calculation of \cite{Caputa:2024sux}.

\subsection{$AdS_3$ subspace of $AdS_5\times S^5$ and its Krylov complexity}

The relevant $AdS_3$ subspace is obtained by writing the $AdS_5$ metric as 
\bea
ds^2&=& R^2\left[-dt^2\cosh^2\rho +d\rho^2+\sinh^2 \rho d\Omega_3^2\right]\cr
&=& R^2\left[-dt^2\cosh^2\rho +d\rho^2+\sinh^2\rho 
\left(\cos^2\theta d\phi^2+d\theta^2 +\sin^2\theta d\psi^2\right)\right]\;,
\eea
and the $AdS_3$ subspace in which the $Sl(2)$ subsector lives (on its boundary) is 
the Equator $\theta=0$, with metric 
\be
ds^2_3=R^2\left[-dt^2\cosh^2\rho +d\rho^2+\sinh^2\rho d\phi^2\right].
\ee
Note that we have an $AdS_3$ {\em only} at $\theta=0$. This subspace is what contained the 
GKP spinning folded string \cite{Gubser:2002tv}.
Another (equivalent) form is given in \cite{Kazakov:2004nh} as in the embedding (hyperboloid) coordinates 
$X_{-1},X_0,X_1,X_2,X_3,X_4$ for $AdS_5$, one fixes $X_3,X_4$ to constants, obtaining $AdS_3$. 

At the boundary $\rho\rightarrow \infty$, we have 
\be
ds_2^2\simeq\frac{R^2e^{2\rho}}{4}(-dt^2+d\phi^2)\;,
\ee
so the $\mathbb{R}_t\times S^1$ of the cylinder ($CFT_2$ on a finite line), in 
$w=t_E+i\phi$ complex Euclideanized coordinates, defined only at $\theta
=0$. The corresponding plane (for $CFT_2$ on the infinite line) in  $z=e^w=e^{t_E+i\phi}=x_2+ix_1$ 
complex coordinates is also only defined for $\theta=0$ in the $S^3$ 
(at its equator). 

The Krylov complexity for this $AdS_3$ subspace is then exactly (\ref{KrylovL}) in global coordinates, 
corresponding to the $CFT_2$ on the cylinder, in $w$ 
coordinates, and (\ref{Krylovinf}) in Poincar\'{e} coordinates, corresponding to the $CFT_2$ on the plane, 
in $z$ coordinates.

\subsection{$Sl(2)$ subsector of ${\cal N}=4$ SYM and its Krylov complexity}

Corresponding to the $AdS_3$ subspace above, we have the $Sl(2)$ subspace of ${\cal N}=4$ SYM, 
defined by Beisert in \cite{Beisert:2003jj}. The generators of the relevant $sl(2)$ subalgebra of the 
$so(4,2)$ algebra of ${\cal N}=4 SYM$ (a $CFT_4$) are:
{\em $sl(2)\times sl(2)$ Wick rotated Lorentz generators within $CFT_4$}, 
for $SO(3,1)\rightarrow SO(2,2)\simeq SO(2,1)\times SO(2,1)\simeq Sl(2)\times Sl(2)$, 
 called ${L^a}_b$ and ${{\tilde{L}}^a\,}_b$, also 4 dimensional momenta $P^a, P^{\dot a}$, 4 dimensional
boosts $K^a, K^{\dot a} $  and $D$ (dilatations in $CFT_4$, associated with the {\em overall} radial direction 
of $AdS_5$). Here $a,\dot a=1,2$.

Then, the $sl(2)$ generators for the symmetry of the $Sl(2)$ subsector are
\bea
J'_+&=&P_1+iP_2\;,\cr
J'_-&=& K_1+iK_2\;,\cr
J'_3&=& \frac{1}{2}(D+\delta D +{L^1}_1+{{\tilde L^1}\,}_1)\;,
\eea
where $\delta D$ is the quantum correction to $D$, satisfying 
\be
[J'_+,J'_-]=-2J'_3\;,\;\;[J'_3,J'_\pm]=\pm J'_\pm.
\ee

As we can see, this $Sl(2)$ is associated, in $AdS_5$, with the {\em overall} radial direction $\rho$, 
the Euclideanized time direction 2, and a spatial direction 1. 

The generators $J'_\pm,J'_3$ are represented in terms of oscillators as 
\be
J'_-=a\;,\;\;J'_+=a^\dagger +a^\dagger a^\dagger a\;,\;\;
J'_3=\frac{1}{2}+a^\dagger a.
\ee

Then,  the {\em Sl(2) sector of ${\cal N}=4$ SYM} is defined by combinations of the objects
(complex scalar and complex derivative)
\be
Z=\Phi_{5+i6}=\Phi_5+i\Phi_6\;,\;\;
D_+=D_{1+i2}=D_1+iD_2\;,
\ee
where $D_+$ {\em can be defined either on the cylinder $w$ or on the plane $z$, as above,}
and the identification with the oscillator states is given by
\be
|n\rangle =\frac{1}{n!}(a^\dagger)^n|0\rangle=\frac{1}{n!}D_+^n|0\rangle\;,
\ee
which corresponds in the ${\cal N}=4$ SYM CFT to operators of given spin  $n=S$ and given dimension,
for instance
\be
{\cal O}_S\sim \Tr[D_+^SZ].
\ee
This is an operator corresponding in a spin chain to only parallel spins.

More generally, we can construct a {\em spin chain} by using the general operator
\be
{\cal O}_{m_1,...,m_L}=\Tr[D_+^{m_1}ZD_+^{m_2}Z...D_+^{m_L}Z]\;,
\ee
with the $D_+^{m_k}$ identified with spins inserted at lattice site $k$, and with spin $S$, dilatation 
charge $\Delta$ and R-charge $L$ given by
\be
S=\sum_k m_k\;,\;\; L=\sum_k 1\;,\;\; \Delta_{g=0}=S+L.
\ee

The one-loop Hamiltonian found by Beisert in \cite{Beisert:2003jj} is in fact the XXX spin 1/2 Heisenberg
Hamiltonian, which is nearest-neighbour {\em in site ($k$) space}. 

Thus, definining, within the subspace of operators
${\cal O}_{m_1,...m_L}$ the operators of given $S$ and $L$, ordered by site (increasing $m_k$, then 
increasing $k$) insertion, it is also 
nearest-neighbour in the Krylov ($|n\rangle$) sense, so can be identified with the Krylov basis, so the 
corresponding Heisenberg $XXX_{1/2}$ Hamiltonian with the Krylov basis Hamiltonian (\ref{HKrylov}). 
This is so, since both the subsector Hamiltonian and the Krylov Hamiltonian are for quantum mechanical 
(0+1 dimensional) finite dimensional systems.

One could then just compute the complexity from the Hamiltonian in this Krylov basis (which we leave for 
further work) but, given 
that for generic $AdS_3/CFT_2$ this was done in \cite{Caputa:2024sux} (in Supplemental Material A), 
there is no need to do it here as well. 

Note that for only two $D_+$ insertions in a chain of length $L$, the operators of 
momentum $n$ that are {\em eigenstates
of the Hamiltonian} (but therefore {\em not Krylov basis states}: diagonal vs. tri-diagonal form) are 
\be
{\cal O}_n^L=2\cos\frac{\pi n}{L+1}\Tr(a_1^\dagger)^2|L\rangle+
\sum_{p=2}^L\cos \frac{\pi n(2p-1)}{L+1}a_1^\dagger a_p^\dagger |L\rangle\;,
\ee
which are the 2-magnon BMN \cite{Berenstein:2002jq} 
operators, of energy and correction to the dilatation operator of 
\be
E=8\sin^2\frac{\pi n}{L+1}\;,\;\;
\delta D =\frac{g^2_{YM}N}{\pi^2}\sin^2\frac{\pi n}{L+1}.
\ee

On the other hand, in the opposite case of large spin $S$, so large $\Delta$, but with fixed $J$, 
the gravity dual corresponding to the operator that is an eigenstate of the Hamiltonian is the folded spinning 
string at the center of $AdS_5$, but restricted to move in $AdS_3$, of Gubser-Klebanov-Polyakov, 
\cite{Gubser:2002tv}, with behaviour 
\be
\Delta-S=\frac{\sqrt{\lambda}}{\pi }\ln \frac{S}{\sqrt{\lambda}}+...\;,\;\; \lambda=g^2_{YM}N.
\ee

In conclusion, the generic $Sl(2)$ subsector state moves in $AdS_3$ only (it is not necessarily a large string, 
but maybe just a particle), and stays inside it, consistent with what we have found for the 
complexity. 

\section{Motion in $AdS_5$ sliced by $AdS_3$ and holographic Krylov complexity}

We now consider a more general case, where in the gravity dual of $AdS_5$ (we ignore the $S^5$, considering
that there is no motion there) we have a motion that still respects the $Sl(2)$ symmetry at every given point, 
but the motion itself is not $Sl(2)$ symmetric. The result of the gravity dual calculation will be then 
considered as a prediction, if we consider that the matching we had in the $AdS_3$ vs. $Sl(2)$ subsector
should extend to the full $Sl(2)$ invariant theory.

\subsection{Motion in $AdS_5$ sliced by $AdS_3$}

That means that we must find a slicing of $AdS_5$ by $AdS_3$ slices, which in the more general 
case of $AdS_{d+2}$ sliced by $AdS_d$ is given by
\be
ds^2_{AdS_{d+2}}=\sinh^2 r d\psi^2+dr^2+\cosh^2 r ds^2_{AdS_d}. 
\ee

\subsubsection{Global coordinates for $AdS_3$}

In the case of $AdS_5$ vs. $AdS_3$ and global coordinates, we have  
\be
ds^2=dr^2+\sinh^2 r d\psi^2+\cosh^2 r (d\rho^2-\cosh^2 \rho dt^2+\sinh^2 \rho d\phi^2)\;,
\ee
so, with the ansatz for timelike geodesics with $r=r(t),\rho=\rho(t)$, the Lagrangian for a 
massive particle is 
\be
S=-m\int\; dt\sqrt{\cosh^2 r (\cosh^2 \rho -\dot \rho^2)-\dot r^2}.\label{globalaction}
\ee
In this case, the proper distance will be found by replacing the solution with $r(t),\rho(t)$ in 
\be
ds^2=dr^2+\cosh^2 r d\rho^2
\ee
and rewriting this as a $d\bar \rho^2$.
%in which case the proper momentum will be $P_R$. 

The canonical momenta associated with the variables $r(t),\rho(t)$ are 
\be
P_\rho =\frac{m\cosh^2 r \dot \rho}{\sqrt{\cosh^2 r (\cosh^2 \rho -\dot \rho^2)-\dot r^2}}\;,\;\;
P_r=\frac{m\dot r}{\sqrt{\cosh^2 r (\cosh^2 \rho -\dot \rho^2)-\dot r^2}}\;,
\ee
and the resulting conserved (time-independent) Hamiltonian is 
\be
H=P_\rho \dot\rho+P_r\dot r-L=m\frac{\cosh^2 r\cosh^2 
\rho}{\sqrt{\cosh^2 r (\cosh^2 \rho -\dot \rho^2)-\dot r^2}}\;,
\ee
which means that, from the constancy of $H$, we can  solve for $\dot r$ as
\bea
\dot r^2&=&\cosh^2 r (\cosh^2\rho -\dot \rho^2)-\frac{m^2\cosh^4 r \cosh^4 \rho}{ H^2}\cr
&=&\cosh^2 r \left[\cosh^2\rho\left(1-\frac{m^2}{H^2}\cosh^2 r \cosh^2 \rho\right)-\dot \rho^2\right]\;.\label{rdot}
\eea
On the other hand,  the equations of motion of the action (\ref{globalaction}), for $\rho$ and $r$, 
respectively, are 
\bea
\frac{d}{dt}\left[\frac{m\cosh^2 r \dot \rho}{\sqrt{\cosh^2 r (\cosh^2 \rho -\dot \rho^2)-\dot r^2}}\right]
&=& -m\frac{\cosh \rho \sinh \rho \cosh^2 r}{\sqrt{\cosh^2 r (\cosh^2 \rho -\dot \rho^2)-\dot r^2}}\cr
\frac{d}{dt}\left[\frac{m\dot r}{\sqrt{\cosh^2 r (\cosh^2 \rho -\dot \rho^2)-\dot r^2}}\right]
&=& -m \frac{\cosh r \sinh r \cosh^2\rho}{\sqrt{\cosh^2 r (\cosh^2 \rho -\dot \rho^2)-\dot r^2}}.
\eea

In the case of $AdS_3$, the initial boundary condition for the massive geodesic was that it starts near 
the boundary (at $\rho_0\rightarrow\infty$), with zero velocity, $\dot \rho_0=0$. 

In the case of motion in $AdS_5$, with $r(t),\rho(t)$, we will assume that (since we want to consider 
$AdS_3$ as a basis for the calculation), together with the boundary condition for $r$ that it also starts near 
the boundary, at large $r$, $r_0\rightarrow\infty$. 

Note that, for $r=r_0\rightarrow \infty, \rho=\rho_0\rightarrow \infty$, the $AdS_5$ (boundary) metric becomes
\be
ds^2\simeq \frac{e^{2r_0}}{4}\left[d\psi^2+\frac{e^{2\rho_0}}{4}(-dt^2+d\phi^2)\right]
=\frac{e^{2r_0+2\rho_0}}{16}\left[4e^{-2\rho_0}d\psi^2-dt^2+d\phi^2\right]\;,
\ee
so the circle $d\psi^2$ becomes of zero length, and can be dropped, so the boundary metric is 
just $\mathbb{R}_t\times S^1$, just the boundary of global $AdS_3$, consistent with a modification of the 
$Sl(2)$ sector described by similar operators. 

The simplest and most interesting case is when 
we also start at zero velocity in $r$, so $\dot r_0=0$, in which case from (\ref{rdot}), 
we get
\be
H=m\cosh r_0\cosh\rho_0\;.
\ee
Then, eliminating $m/H$, we get 
\be
\dot r=\cosh r \sqrt{\cosh^2\rho\left(1-\frac{\cosh^2\rho \cosh^2 r}{\cosh^2\rho_0\cosh^2 r_0}\right)
-\dot \rho^2}\;,
\ee
but we will not use this for now.

Substituting 
\be
m\frac{\cosh^2 r}{\sqrt{\cosh^2 r (\cosh^2 \rho -\dot \rho^2)-\dot r^2}}=\frac{H}{\cosh^2\rho}
\ee
in the $\rho$ equation of motion, and thus getting rid of $\dot r$, we obtain the constant Hamiltonian
 $H$ on both sides of the equation,
so it cancels out, and we remain with the simple equation
\be
\frac{d}{dt}\left[\frac{\dot \rho}{\cosh^2\rho}\right]=-\tanh \rho.
\ee

At general $\rho$, we write the equation as
\be
\frac{d^2}{dt^2} \tanh \rho +\tanh \rho=0\Rightarrow \tanh \rho=A' \cos t+B' \sin t=A\cos(t-t_0)\;,
\ee
or (imposing that at $\rho=\rho_0$ we have $\dot \rho_0=0$), 
\be
\tanh \rho =\tanh \rho_0 \cos (t-t_0).
\ee

Note that at large $\rho$, so for small times ($t$ near $t_0$, so $\rho$ near $\rho_0\rightarrow \infty$), we 
get 
\be
\rho\simeq -\frac{1}{2}\log \left[e^{-2\rho_0}+\frac{(t-t_0)^2}{4}\right].
\ee

To find $r(t)$, we first write the conservation of $H/m$ as 
\be
\frac{H^2}{m^2}\frac{\dot r^2}{\cosh^2 r}=\frac{H^2}{m^2}\cosh^2\rho -\cosh^2 r \cosh^4 \rho-\frac{H^2}{m^2}
\dot \rho^2\;,
\ee
then, in the $\dot r_0=0$ case, we 
replace $H/m$ with $\cosh r_0\cosh\rho_0$ on the right-hand side, and then substitute the solution for 
$\rho(t)$ and so also for $\dot \rho (t)$ on the right-hand side, to obtain 
\be
\frac{H^2}{m^2}\frac{\dot r^2}{\cosh^2 r}=\frac{1}{(1-\tanh^2 \rho_0 \cos^2(t-t_0))^2}
\left[\cosh^2 r_0-\cosh^2 r\right]\;,\label{dotr}
\ee
and then taking the square root and separating the $dr$ and $dt$ integrals gives 
\be
\frac{H}{m}\int_{r_0}^r \frac{dr}{\cosh r\sqrt{\cosh^2 r_0-\cosh^2r}}=
\int_{t_0}^t \frac{dt}{1-\tanh^2 \rho_0 \cos^2 (t-t_0)}\;,
\ee
and doing the integrals on both sides, 
\bea
&&\frac{H}{m}\left.\frac{1}{\cosh r_0}\left(\arcsin \frac{\tanh r}{\tanh r_0}\right)\right|_{r_0}^r
= -\left.\cosh \rho_0\arctan \left[\frac{\cot (t-t_0)}{\cosh \rho_0}\right]\right|_{t_0}^t\Rightarrow\cr
&&r(t)=\tanh^{-1}\left\{\tanh r_0\sin\left[-\arctan \left[\frac{\cot (t-t_0)}{\cosh \rho_0}\right]
+\frac{3\pi}{4}\right]\right\}  \;.\label{roftsol}
\eea

At this point it seems from (\ref{dotr}) that $r=r_0=$constant is a solution,
yet, the equation of motion for $r$, which can be rewritten, by substituting $H$ on both sides of the 
equation, and cancelling this constant value (as for the equation of motion for $\rho$) gives
\be
\frac{d}{dt}\left[\frac{\dot r}{\cosh^2 r \cosh^2\rho}\right]=-\tanh r\;,
\ee
which clearly doesn't admit the $r=r_0$ solution (the left-hand side is zero, but the right-hand side is not). 
If $\dot r_0=0$, then we
find $\ddot r_0=-\cosh^2 r_0 \tanh r_0 \cosh^2\rho_0$, very large.
But there is no need to go to this, we can already see in the solution in (\ref{roftsol}) that $r=r_0$ is not  
possible: in fact, the solution is unique, without any free constants. 

We can also choose the more general boundary condition for $r$, namely a nonzero velocity, 
$\dot r_0\neq 0$, in which case we just keep $H/m$ as it is, and obtain instead 
of (\ref{dotr}) that
\be
\frac{H^2}{m^2}\frac{\dot r^2}{\cosh^2 r}=\frac{1}{(1-\tanh^2 \rho_0 \cos^2(t-t_0))^2}
\left[\frac{H^2}{m^2\cosh^2\rho_0}-\cosh^2 r\right]\;,\label{dotr2}
\ee
so that
\be
\frac{H}{m}\int_{r_0}^r \frac{dr}{\cosh r\sqrt{\frac{H^2}{m^2\cosh^2\rho_0}-\cosh^2r}}=
\int_{t_0}^t \frac{dt}{1-\tanh^2 \rho_0 \cos^2 (t-t_0)}\;,
\ee
which integrates to
\bea
&&\cosh\rho_0\left.\left(\arcsin \frac{m\cosh\rho_0\tanh r}{\sqrt{H^2-m^2\cosh^2\rho_0}}\right)\right|_{r_0}^r\cr
&=& -\left.\cosh \rho_0\arctan \left[\sqrt{1-\tanh^2\rho_0}\cot (t-t_0)\right]\right|_{t_0}^t\;.
\eea
or 
\bea
&&\left.\left(\arcsin \frac{m\cosh\rho_0\tanh r}{\sqrt{H^2-m^2\cosh^2\rho_0}}\right)\right|_{r_0}^r
= -\left.\arctan \left[\frac{\cot (t-t_0)}{\cosh \rho_0}\right]\right|_{t_0}^t\Rightarrow\cr
r(t)&=& \tanh^{-1}\left\{\frac{\sqrt{H^2-m^2\cosh^2\rho_0}}{m\cosh \rho_0}\sin\left[-\arctan \left(\frac{\cot (t-t_0)}
{\cosh \rho_0}\right)+\frac{\pi}{2}\right.\right.\cr
&&\left.\left.+\arcsin\frac{m\cosh \rho_0 \tanh r_0}{\sqrt{H^2-m^2\cosh^2\rho_0}}\right]
\right\}\;.
\eea

\begin{figure}
    \centering
    \includegraphics[width=0.7\linewidth]{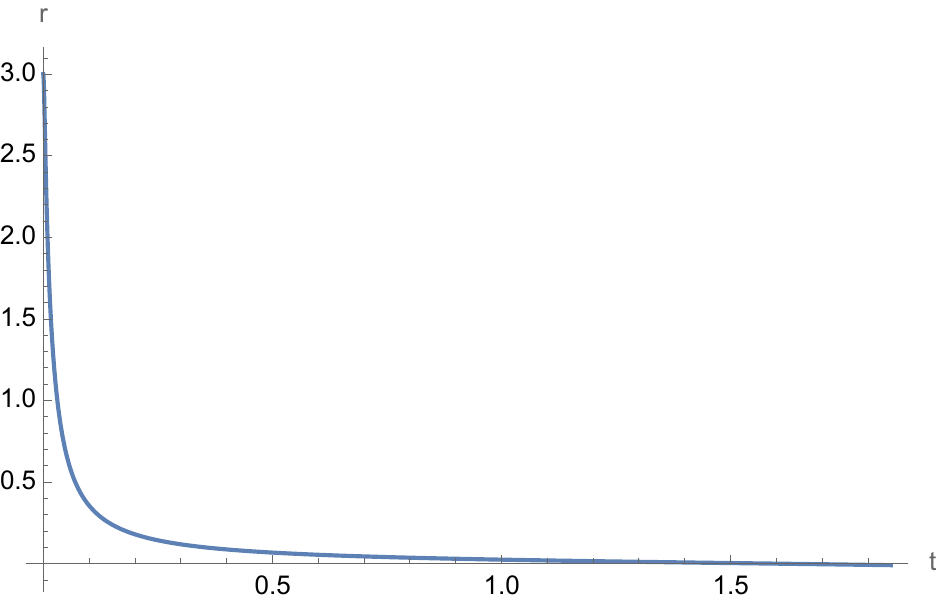}
    \caption{The evolution of $r(t)$ for $AdS_3$ in global coordinates that follows from \eqref{dotr2}. The initial conditions are set as $\dot r_0=0.1$, 
    $\rho_0=4, r_0=3$, together with $t_0=0$. }
    \label{fig_r_global_ads}
\end{figure}

An example of evolution of $r(t)$ is in Fig.\ref{fig_r_global_ads}.

\subsubsection{Poincar\'{e} coordinates for $AdS_3$}

The slicing of $AdS_5$ by $AdS_3$ is  independent on the metric of $AdS_3$, so we can use 
the Poincar\'{e} patch of $AdS_3$, described, like in \cite{Caputa:2024sux}, in coordinates
\be
ds^2_3=d\rho^2+e^{2\rho}(-dt^2+dx^2)\;,
\ee
and corresponding to the $CFT_2$ on an infinite line. Indeed, the boundary of $AdS_5$ at $r=r_0\rightarrow
\infty$ and $\rho=\rho_0\rightarrow \infty$ is now 
\be
ds^2_5=\frac{e^{2r_0+2\rho_0}}{4}\left[e^{-2\rho_0}d\psi^2-dt^2+dx^2\right]\;,
\ee
so just 2-dimensional Minkowski space, $\mathbb{R}_t\times \mathbb{R}_x$, 
corresponding to a $CFT_2$ on an 
infinite line, meaning that we are calculating the complexity for a $CFT_2$ on an infinite line.

The calculation in Poincar\'{e} $AdS_3$ follows the same one in the global case. With the $r=r(t), \rho
=\rho(t)$ ansatz, we have the massive particle action 
\be
S=-m\int dt \sqrt{\cosh^2 r (e^{2\rho}-\dot \rho^2)-\dot r^2}\;,
\ee
giving the canonical momenta 
\be
P_\rho =\frac{m\cosh^2 r \dot \rho}{\sqrt{\cosh^2 r (e^{2 \rho} -\dot \rho^2)-\dot r^2}}\;,\;\;
P_r=\frac{m\dot r}{\sqrt{\cosh^2 r (e^{2 \rho} -\dot \rho^2)-\dot r^2}}\;,
\ee
and the conserved Hamiltonian is 
\be
H=P_\rho \dot\rho+P_r\dot r-L=m\frac{\cosh^2 r e^{2\rho}}{\sqrt{\cosh^2 r (e^{2 \rho} -\dot \rho^2)-\dot r^2}}\;,
\ee
so that, solving for $\dot r$, we obtain
\be
\dot r^2=\cosh^2 r (e^{2\rho} -\dot \rho^2)-\frac{m^2\cosh^4 r e^{4 \rho}}{ H^2}
=\cosh^2 r \left[e^{2\rho}\left(1-\frac{m^2}{H^2}\cosh^2 r e^{2 \rho}\right)-\dot \rho^2\right]\;.\label{rdot}
\ee
The equations of motion for $\rho$ and $r$ coming from the massive particle action are 
\bea
\frac{d}{dt}\left[\frac{m\cosh^2 r \dot \rho}{\sqrt{\cosh^2 r (e^{2 \rho} -\dot \rho^2)-\dot r^2}}\right]
&=& -m\frac{e^{2\rho} \cosh^2 r}{\sqrt{\cosh^2 r (e^{2 \rho} -\dot \rho^2)-\dot r^2}}\cr
\frac{d}{dt}\left[\frac{m\dot r}{\sqrt{\cosh^2 r (e^{2 \rho} -\dot \rho^2)-\dot r^2}}\right]
&=& -m \frac{\cosh r \sinh r (e^{2\rho}-\dot{\rho}^2)}{\sqrt{\cosh^2 r (e^{2 \rho} -\dot \rho^2)-\dot r^2}}.
\eea

Substituting the constant Hamiltonian on both sides of the first equation, it cancels, and we obtain simply
(using the same boundary conditions)
\be
\frac{d^2}{dt^2}e^{-2\rho}=+2\Rightarrow \rho(t)=-\frac{1}{2}\log\left[e^{-2\rho_0}+(t-t_0)^2\right].
\ee

To find $r(t)$, we first write the conservation of $H/m$ as 
\be
\frac{H^2}{m^2}\frac{\dot r^2}{\cosh^2 r}=\frac{H^2}{m^2}e^{2\rho} -\cosh^2 r e^{4 \rho}-\frac{H^2}{m^2}
\dot \rho^2\;,
\ee
then replace $H/m$ with $\cosh r_0 e^{\rho_0}$ on the right-hand side, and then substitute the solution for 
$\rho(t)$ and so also for $\dot \rho (t)$ on the right-hand side, to obtain 
\be
\frac{H^2}{m^2}\frac{\dot r^2}{\cosh^2 r}=\frac{e^{4\rho_0}}{(1+e^{2 \rho_0} (t-t_0))^2}
\left[\cosh^2 r_0-\cosh^2 r\right]\;,
\ee
so taking the square root and separating the $dr$ and $dt$ integrals gives 
\be
\frac{H}{m}\int_{r_0}^r \frac{dr}{\cosh r\sqrt{\cosh^2 r_0-\cosh^2r}}=
\int_{t_0}^t \frac{dt}{e^{-2\rho_0}+  (t-t_0)^2}\;,
\ee
or, doing the integrals on both sides, 
\bea
&&\frac{H}{m}\left.\frac{1}{\cosh r_0}\left(\arcsin \frac{\tanh r}{\tanh r_0}\right)\right|_{r_0}^r
= \left.e^{\rho_0}\arctan \left[e^{\rho_0} (t-t_0)\right]\right|_{t_0}^t\Rightarrow\cr
r(t)&=&\tanh^{-1}\left\{\tanh r_0 \sin\left[\arctan \left[e^{\rho_0} (t-t_0)\right]+\frac{\pi}{4}\right]\right\}\;.
\eea

Again for the boundary condition $\dot r_0=0$ we seem to have the solution $r=r_0=$constant, 
but it is actually not a solution of the equation for 
$r$, which is (substituting $H$ on both sides of the equation, and then cancelling it)
\be
\frac{d}{dt}\left[\frac{\dot r e^{-2\rho}}{\cosh^2 r}\right]=-\tanh r (1-\dot{\rho}^2 e^{-2 \rho})\;,
\ee
so for $\dot r_0=0$ we would obtain the initial acceleration $\ddot r_0=-\cosh^2 r_0\tanh r_0 e^{2\rho_0}$, 
very large. 

For the more general boundary condition, $\dot r_0\neq 0$, we leave 
$H/m$ as it is, and obtain 
\be
\label{e4.39}
\frac{H^2}{m^2}\frac{\dot r^2}{\cosh^2 r}=\frac{e^{4\rho_0}}{(1+e^{2 \rho_0} (t-t_0))^2}
\left[\frac{H^2}{m^2}e^{-2\rho_0}-\cosh^2 r\right]\;,
\ee
so that
\be
\frac{H}{m}\int_{r_0}^r \frac{dr}{\cosh r\sqrt{\frac{H^2}{m^2}e^{-2\rho_0}-\cosh^2r}}=
\int_{t_0}^t \frac{dt}{e^{-2\rho_0}+  (t-t_0)^2}\;,
\ee
and so
\bea
&&e^{\rho_0}\left.\left(\arcsin \frac{m e^{\rho_0}\tanh r}{\sqrt{H^2-m^2 e^{2\rho_0}}}\right)\right|_{r_0}^r
= \left.e^{\rho_0}\arctan \left[e^{\rho_0} (t-t_0)\right]\right|_{t_0}^t\Rightarrow\cr
r(t)&=&\tanh ^{-1}\left\{\frac{\sqrt{H^2-m^2 e^{2\rho_0}}}{me^{\rho_0}}\sin\left[\arctan\left(e^{\rho_0}(t-t_0)\right)
+\arcsin\frac{me^{\rho_0}\tanh r_0}{\sqrt{H^2-m^2e^{2\rho_0}}}\right]\right\}\;.\cr
&&
\eea

\begin{figure}
    \centering
    \includegraphics[width=0.7\linewidth]{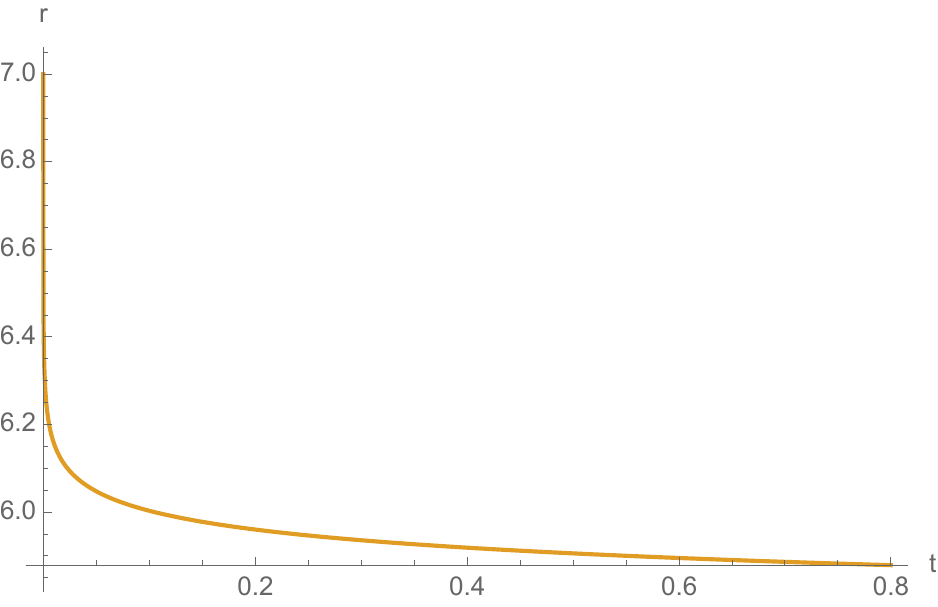}
    \caption{We evaluate $r(t)$ for $AdS_3$ in Poincar\'{e} coordinates following \eqref{e4.39}. The initial conditions are set to be $\dot r_0=0.01$, $\rho_0=8, r_0=7$.}
    \label{fig_r_Poincare}
\end{figure}

An example of evolution of $r(t)$ is in Fig.\ref{fig_r_Poincare}.

\subsection{Proper momentum as derivative of Krylov complexity in ${\cal N}=4$ SYM}

We now calculate the proper momentum $P_{\bar\rho}$ which, by the proposed formula, gives 
the time derivative of the Krylov complexity, $\dot C(t)$.

\subsubsection{Global coordinates for $AdS_3$}

In global coordinates, we define the proper distance by 
\be
d\bar \rho^2=dr^2+\cosh^2 rd\rho^2\;,
\ee
and since the canonical momenta re 
\be
P_\rho=\frac{\d L}{\d \dot \rho}=\frac{H\dot \rho}{\cosh^2\rho}\;,\;\;
P_r=\frac{\d L}{\d \dot r}=\frac{H \dot r}{\cosh^2 \rho \cosh^2 r}\;,
\ee
the proper momentum is 
\bea
\label{e4.44}
P_{\bar \rho}&=&\frac{P_r}{\sqrt{1+\cosh^2 r \rho'(r)^2}}+\frac{P_\rho}{\sqrt{r'^2 +\cosh^2 r}}\cr
&=& \frac{H\dot r}{\cosh^2 \rho \cosh^2 r}\frac{1}{\sqrt{1+\cosh^2 r \rho'^2(r)}}+\frac{H \dot \rho \rho'(r)}{\cosh^2 
\rho \sqrt{1+\cosh^2 r \rho'^2(r)}}\cr
&=& \frac{H}{\cosh^2 \rho\cosh^2 r}\dot \rho\sqrt{\cosh^2 r+\frac{\dot r^2}{\dot \rho^2}}\cr
&=&\frac{H}{\cosh \rho \cosh r}\sqrt{1-\frac{m^2\cosh^2\rho\cosh^2 r}{H^2}}\;,
\eea
where we have substituted $\dot \rho,\dot r$ in terms of just $r,\rho$ from their solutions. 

If $\dot r_0=0$, we substitute $H/m$ in terms of $r_0,\rho_0$ and obtain the simpler form
\be
P_{\bar\rho}=\frac{H}{\cosh \rho \cosh r}\sqrt{1-\frac{\cosh^2\rho\cosh^2 r}{\cosh^2 \rho_0\cosh^2 r_0}}\;.
\ee

\begin{figure}
    \centering
    \includegraphics[width=0.7\linewidth]{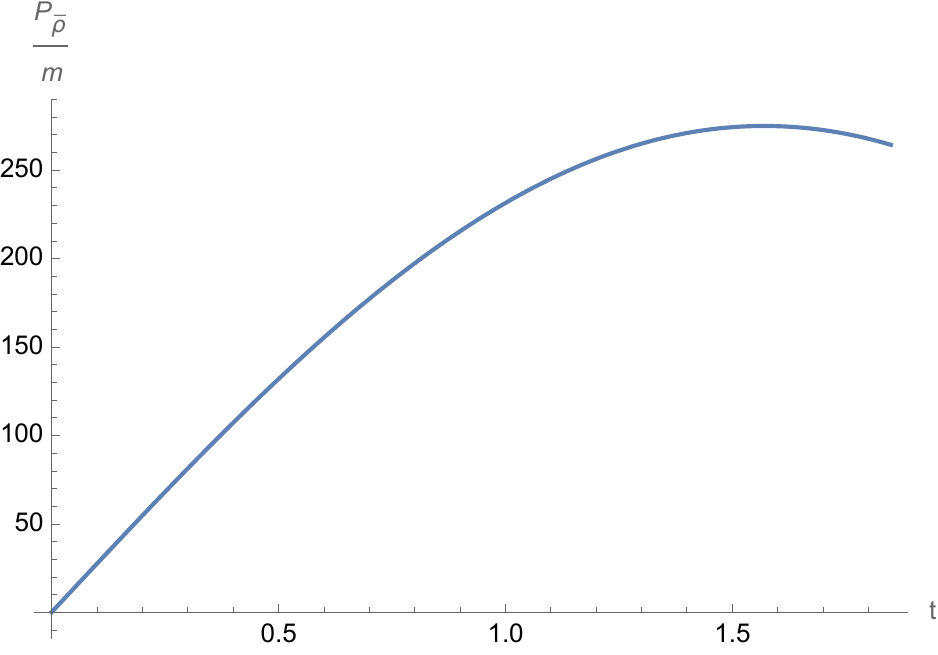}
    \caption{The evolution of the proper momentum $P_{\bar\rho}(t)$ 
    for $AdS_3$ in global coordinates has been carried out following \eqref{e4.44}. The initial conditions are set to be $\dot r_0=0.1$, 
    $\rho_0=4, r_0=3$, $t_0=0$. }
    \label{fig_P_global_ads}
\end{figure}

An example of evolution of the proper momentum in this global case is in Fig.\ref{fig_P_global_ads}.

Note that if we would also have $r=r_0$, that would mean
\be
P_{\bar \rho}=\frac{H}{\cosh \rho \cosh r_0}\sqrt{1-\frac{\cosh^2\rho}{\cosh^2\rho_0}}=m\tanh \rho_0 
\sin (t-t_0)\;,
\ee
which is indeed the result for $AdS_3$ in global coordinates but, as we saw, this is not a good solution.

\subsubsection{Poincar\'{e} coordinates for $AdS_3$}

In the Poincar\'{e} case, we still have the same proper distance definition,
\be
d\bar \rho^2=dr^2+\cosh^2 rd\rho^2\;,
\ee
and the canonical momenta are now
\be
P_\rho=\frac{\d L}{\d \dot \rho}=\frac{H\dot \rho}{e^{2\rho}}\;,\;\;
P_r=\frac{\d L}{\d \dot r}=\frac{H \dot r}{e^{2 \rho} \cosh^2 r}\;,
\ee
so the proper momentum is
\bea
\label{e4.49}
P_{\bar \rho}&=&\frac{P_r}{\sqrt{1+\cosh^2 r \rho'(r)^2}}+\frac{P_\rho}{\sqrt{r'^2 +\cosh^2 r}}\cr
&=& \frac{H\dot r}{e^{2 \rho} \cosh^2 r}\frac{1}{\sqrt{1+\cosh^2 r \rho'^2(r)}}+\frac{H \dot \rho \rho'(r)}{e^{2 
\rho} \sqrt{1+\cosh^2 r \rho'^2(r)}}\cr
&=& \frac{H}{e^{2 \rho}\cosh^2 r}\dot \rho\sqrt{\cosh^2 r+\frac{\dot r^2}{\dot \rho^2}}\cr
&=&\frac{H}{e^\rho \cosh r}\sqrt{1-\frac{m^2}{H^2}e^{2\rho}\cosh^2 r}\;,
\eea
where we have substituted $\dot \rho,\dot r$ in terms of $\rho,r$.

If $\dot r_0=0$, we substitute $H/m$ in terms of $r_0,\rho_0$ and obtain
\be
P_{\bar\rho}=\frac{H}{e^\rho \cosh r}\sqrt{1-\frac{e^{2\rho}\cosh^2 r}{e^{2 \rho_0}\cosh^2 r_0}}\;.
\ee

\begin{figure}
    \centering
    \includegraphics[width=0.7\linewidth]{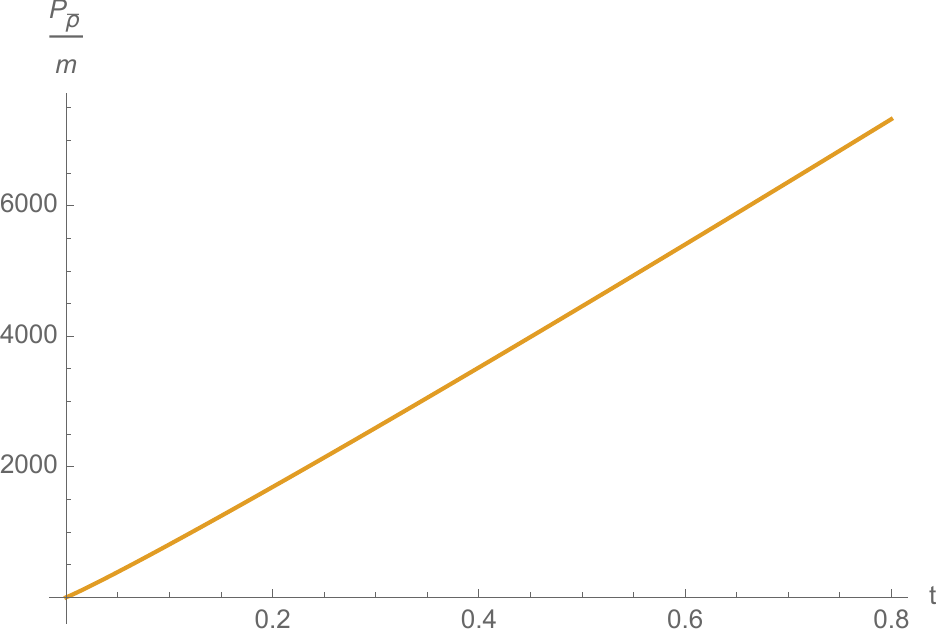}
    \caption{The proper momentum for $AdS_3$ in Poincar\'{e} coordinates has been estimated using \eqref{e4.49}. The initial conditions are set to be $\dot r_0=0.01$, $\rho_0=8, r_0=7$. }
    \label{fig_P_Poincare}
\end{figure}

An example of momentum dependence is in Fig.\ref{fig_P_Poincare}.

If we would have also $r=r_0$, then we would obtain 
\be
P_{\bar \rho}=\frac{H}{ \cosh r_0}e^{-\rho}\sqrt{1-e^{2(\rho-\rho_0)}}=m e^{\rho_0}(t-t_0)\;,
\ee
which is indeed the result for $AdS_3$ in Poincar\'{e} coordinates but, as we saw, this is not a good solution.

\subsubsection{Field theory interpretation}

As we saw, the dual to the $AdS_3$ subspace is the $Sl(2)$ sector of ${\cal N}=4$ SYM, and the 
complexity is considered within it, in that case. The subsector is obtained by acting with 
$D_+=D_1+iD_2$ on $Z=\Phi_5+i\Phi_6$, where $x_1+ix_2=z=e^{w}=e^{t_E+i\phi}$, and it has $Sl(2)$ 
symmetry. 

In our case, however, there is a sort of instantaneous (fixed time)
$Sl(2)$ symmetry, due to the $AdS_3$ slicing, and 
resulting instantaneous $AdS_3$ symmetry. Moreover, as we saw, the boundary of $AdS_5$ {\em on 
our $(r(t),\rho(t))$ motion } is still $\mathbb{R}_t\times S^1$, like for $AdS_3$.

One way to describe this is that $\phi$, which was a coordinate 
along an Equator of the $S^3$ in the ${\cal N}=4$ SYM on the $\mathbb{R}_t\times S^3$ cylinder, 
now becomes a coordinate along an {\em instantaneous Equator} defined by the motion along the 
$S^3$ induced by the motion in $AdS_5$, described by $(\rho(t),r(t))$, so we have a sort of $\phi(t)$, and 
corresponding $\d\phi\equiv \d/\d \phi(t)$ derivative. 

Therefore we can say that the $Sl(2)$ sector now varies with time, but stays essentially $Sl(2)$. 
The details of this construction we will leave for further work.

In support of this picture, we saw that the proper momentum, giving the derivative of the Krylov complexity, 
has essentially the same behaviour as in the pure $AdS_3$ case: in global coordinates, corresponding to 
$CFT_2$ on a circle, it has a sinusoidal-like behaviour, due to the finiteness of the Hilbert space, 
while in Poincar\'{e} coordinates, corresponding to $CFT_2$ on the infinite line, it has a linear-like 
behaviour, increasing without bound due to the non-finiteness of the Hilbert space. 

\section{Conclusions}

In this paper, we considered the holographic complexity of ${\cal N}=4$ SYM, extending the 
formula $\dot C(t)=-P_{\bar\rho}/\epsilon$, where $P_{\bar\rho}$ is the proper momentum, defined 
when the evolution from the UV of the gravity dual is written like $ds^2=d\bar\rho^2$ on the motion. 

We took advantage of the fact that $AdS_5$ has a submanifold $AdS_3$ situated at $\theta=0$, 
dual to the $Sl(2)$ sector of ${\cal N}=4$ SYM, and of the fact that we can slice $AdS_5$ by $AdS_3$
slices. The $AdS_3$ case was exactly proven previously, so we just imported it here, and this allowed us 
to predict the behaviour in the general case, as dual to a modified $Sl(2)$ sector. 

There are many issues left for further work. One was the actual calculation of complexity in the $Sl(2)$ 
sector, without using the general $CFT_2$ methods, but rather in a quantum mechanical spin 
chain system (as opposed to a $CFT_2$). Another was the better definition of the deformed $Sl(2)$ sector
giving the holographic result. And, of course, to extend the methods defined here to other cases. 

\section*{Acknowledgments}
We would like to thank Carlos Nunez for participation at the early stages of this project, without whom 
this paper would not have been possible.
The work of HN is supported in part by  CNPq grant 304583/2023-5 and FAPESP grant 2019/21281-4.
HN would also like to thank the ICTP-SAIFR for their support through FAPESP grant 2021/14335-0. DR would like to acknowledge the Mathematical Research Impact Centric Support (MATRICS) grant (MTR/2023/000005) received from ANRF, India.

\bibliographystyle{utphys}
\bibliography{KrylovN4SYM}

\providecommand{\href}[2]{#2}\begingroup\raggedright\begin{thebibliography}{10}

\bibitem{Nielsen:2005mkt}
M.~A. Nielsen, ``{A geometric approach to quantum circuit lower bounds},''
  \href{http://dx.doi.org/10.26421/QIC6.3-2}{{\em Quant. Inf. Comput.} {\bf 6}
  (2006) no.~3, 213--262}, \href{http://arxiv.org/abs/quant-ph/0502070}{{\tt
  arXiv:quant-ph/0502070}}.

\bibitem{Stanford:2014jda}
D.~Stanford and L.~Susskind, ``{Complexity and Shock Wave Geometries},''
  \href{http://dx.doi.org/10.1103/PhysRevD.90.126007}{{\em Phys. Rev. D} {\bf
  90} (2014) no.~12, 126007}, \href{http://arxiv.org/abs/1406.2678}{{\tt
  arXiv:1406.2678 [hep-th]}}.

\bibitem{Brown:2015bva}
A.~R. Brown, D.~A. Roberts, L.~Susskind, B.~Swingle, and Y.~Zhao,
  ``{Holographic Complexity Equals Bulk Action?},''
  \href{http://dx.doi.org/10.1103/PhysRevLett.116.191301}{{\em Phys. Rev.
  Lett.} {\bf 116} (2016) no.~19, 191301},
  \href{http://arxiv.org/abs/1509.07876}{{\tt arXiv:1509.07876 [hep-th]}}.

\bibitem{Belin:2021bga}
A.~Belin, R.~C. Myers, S.-M. Ruan, G.~S\'arosi, and A.~J. Speranza, ``{Does
  Complexity Equal Anything?},''
  \href{http://dx.doi.org/10.1103/PhysRevLett.128.081602}{{\em Phys. Rev.
  Lett.} {\bf 128} (2022) no.~8, 081602},
  \href{http://arxiv.org/abs/2111.02429}{{\tt arXiv:2111.02429 [hep-th]}}.

\bibitem{Jefferson:2017sdb}
R.~Jefferson and R.~C. Myers, ``{Circuit complexity in quantum field theory},''
  \href{http://dx.doi.org/10.1007/JHEP10(2017)107}{{\em JHEP} {\bf 10} (2017)
  107}, \href{http://arxiv.org/abs/1707.08570}{{\tt arXiv:1707.08570
  [hep-th]}}.

\bibitem{Parker:2018yvk}
D.~E. Parker, X.~Cao, A.~Avdoshkin, T.~Scaffidi, and E.~Altman, ``{A Universal
  Operator Growth Hypothesis},''
  \href{http://dx.doi.org/10.1103/PhysRevX.9.041017}{{\em Phys. Rev. X} {\bf 9}
  (2019) no.~4, 041017}, \href{http://arxiv.org/abs/1812.08657}{{\tt
  arXiv:1812.08657 [cond-mat.stat-mech]}}.

\bibitem{Nandy:2024evd}
P.~Nandy, A.~S. Matsoukas-Roubeas, P.~Mart{\'\i}nez-Azcona, A.~Dymarsky, and
  A.~del Campo, ``{Quantum dynamics in Krylov space: Methods and
  applications},'' \href{http://dx.doi.org/10.1016/j.physrep.2025.05.001}{{\em
  Phys. Rept.} {\bf 1125-1128} (2025)  1--82},
  \href{http://arxiv.org/abs/2405.09628}{{\tt arXiv:2405.09628 [quant-ph]}}.

\bibitem{Rabinovici:2025otw}
E.~Rabinovici, A.~S{\'a}nchez-Garrido, R.~Shir, and J.~Sonner, ``{Krylov
  Complexity},'' \href{http://arxiv.org/abs/2507.06286}{{\tt arXiv:2507.06286
  [hep-th]}}.

\bibitem{Baiguera:2025dkc}
S.~Baiguera, V.~Balasubramanian, P.~Caputa, S.~Chapman, J.~Haferkamp, M.~P.
  Heller, and N.~Y. Halpern, ``{Quantum complexity in gravity, quantum field
  theory, and quantum information science},''
  \href{http://arxiv.org/abs/2503.10753}{{\tt arXiv:2503.10753 [hep-th]}}.

\bibitem{Balasubramanian:2022tpr}
V.~Balasubramanian, P.~Caputa, J.~M. Magan, and Q.~Wu, ``{Quantum chaos and the
  complexity of spread of states},''
  \href{http://dx.doi.org/10.1103/PhysRevD.106.046007}{{\em Phys. Rev. D} {\bf
  106} (2022) no.~4, 046007}, \href{http://arxiv.org/abs/2202.06957}{{\tt
  arXiv:2202.06957 [hep-th]}}.

\bibitem{Rabinovici:2023yex}
E.~Rabinovici, A.~S{\'a}nchez-Garrido, R.~Shir, and J.~Sonner, ``{A bulk
  manifestation of Krylov complexity},''
  \href{http://dx.doi.org/10.1007/JHEP08(2023)213}{{\em JHEP} {\bf 08} (2023)
  213}, \href{http://arxiv.org/abs/2305.04355}{{\tt arXiv:2305.04355
  [hep-th]}}.

\bibitem{Ambrosini:2024sre}
M.~Ambrosini, E.~Rabinovici, A.~S{\'a}nchez-Garrido, R.~Shir, and J.~Sonner,
  ``{Operator K-complexity in DSSYK: Krylov complexity equals bulk length},''
  \href{http://dx.doi.org/10.1007/JHEP08(2025)059}{{\em JHEP} {\bf 08} (2025)
  059}, \href{http://arxiv.org/abs/2412.15318}{{\tt arXiv:2412.15318
  [hep-th]}}.

\bibitem{Caputa:2024sux}
P.~Caputa, B.~Chen, R.~W. McDonald, J.~Sim{\'o}n, and B.~Strittmatter,
  ``{Spread Complexity Rate as Proper Momentum},''
  \href{http://arxiv.org/abs/2410.23334}{{\tt arXiv:2410.23334 [hep-th]}}.

\bibitem{Barbon:2020uux}
J.~L.~F. Barbon, J.~Martin-Garcia, and M.~Sasieta, ``{A Generalized
  Momentum/Complexity Correspondence},''
  \href{http://dx.doi.org/10.1007/JHEP04(2021)250}{{\em JHEP} {\bf 04} (2021)
  250}, \href{http://arxiv.org/abs/2012.02603}{{\tt arXiv:2012.02603
  [hep-th]}}.

\bibitem{Das:2024tnw}
R.~N. Das, S.~Demulder, J.~Erdmenger, and C.~Northe, ``{Spread complexity for
  the planar limit of holography},''
  \href{http://dx.doi.org/10.1007/JHEP06(2025)166}{{\em JHEP} {\bf 06} (2025)
  166}, \href{http://arxiv.org/abs/2412.09673}{{\tt arXiv:2412.09673
  [hep-th]}}.

\bibitem{Anabalon:2021tua}
A.~Anabalon and S.~F. Ross, ``{Supersymmetric solitons and a degeneracy of
  solutions in AdS/CFT},''
  \href{http://dx.doi.org/10.1007/JHEP07(2021)015}{{\em JHEP} {\bf 07} (2021)
  015}, \href{http://arxiv.org/abs/2104.14572}{{\tt arXiv:2104.14572
  [hep-th]}}.

\bibitem{Anabalon:2024che}
A.~Anabal{\'o}n, H.~Nastase, and M.~Oyarzo, ``{Supersymmetric AdS solitons and
  the interconnection of different vacua of $ \mathcal{N} $ = 4 Super
  Yang-Mills},'' \href{http://dx.doi.org/10.1007/JHEP05(2024)217}{{\em JHEP}
  {\bf 05} (2024)  217}, \href{http://arxiv.org/abs/2402.18482}{{\tt
  arXiv:2402.18482 [hep-th]}}.

\bibitem{Horowitz:1998ha}
G.~T. Horowitz and R.~C. Myers, ``{The AdS / CFT correspondence and a new
  positive energy conjecture for general relativity},''
  \href{http://dx.doi.org/10.1103/PhysRevD.59.026005}{{\em Phys. Rev. D} {\bf
  59} (1998)  026005}, \href{http://arxiv.org/abs/hep-th/9808079}{{\tt
  arXiv:hep-th/9808079}}.

\bibitem{Lozano:2019emq}
Y.~Lozano, N.~T. Macpherson, C.~Nunez, and A.~Ramirez, ``{AdS$_3$ solutions in
  Massive IIA with small $\mathcal{N}=(4,0)$ supersymmetry},''
  \href{http://dx.doi.org/10.1007/JHEP01(2020)129}{{\em JHEP} {\bf 01} (2020)
  129}, \href{http://arxiv.org/abs/1908.09851}{{\tt arXiv:1908.09851
  [hep-th]}}.

\bibitem{Lozano:2019jza}
Y.~Lozano, N.~T. Macpherson, C.~Nunez, and A.~Ramirez, ``{1/4 BPS solutions and
  the AdS$_3$/CFT$_2$ correspondence},''
  \href{http://dx.doi.org/10.1103/PhysRevD.101.026014}{{\em Phys. Rev. D} {\bf
  101} (2020) no.~2, 026014}, \href{http://arxiv.org/abs/1909.09636}{{\tt
  arXiv:1909.09636 [hep-th]}}.

\bibitem{Lozano:2019zvg}
Y.~Lozano, N.~T. Macpherson, C.~Nunez, and A.~Ramirez, ``{Two dimensional
  ${\cal N}=(0,4)$ quivers dual to AdS$_3$ solutions in massive IIA},''
  \href{http://dx.doi.org/10.1007/JHEP01(2020)140}{{\em JHEP} {\bf 01} (2020)
  140}, \href{http://arxiv.org/abs/1909.10510}{{\tt arXiv:1909.10510
  [hep-th]}}.

\bibitem{Lozano:2019ywa}
Y.~Lozano, N.~T. Macpherson, C.~Nunez, and A.~Ramirez, ``{AdS$_3$ solutions in
  massive IIA, defect CFTs and T-duality},''
  \href{http://dx.doi.org/10.1007/JHEP12(2019)013}{{\em JHEP} {\bf 12} (2019)
  013}, \href{http://arxiv.org/abs/1909.11669}{{\tt arXiv:1909.11669
  [hep-th]}}.

\bibitem{Lozano:2020bxo}
Y.~Lozano, C.~Nunez, A.~Ramirez, and S.~Speziali, ``{$M$-strings and AdS$_3$
  solutions to M-theory with small $\mathcal{N}=(0,4)$ supersymmetry},''
  \href{http://dx.doi.org/10.1007/JHEP08(2020)118}{{\em JHEP} {\bf 08} (2020)
  118}, \href{http://arxiv.org/abs/2005.06561}{{\tt arXiv:2005.06561
  [hep-th]}}.

\bibitem{Lozano:2020txg}
Y.~Lozano, C.~Nunez, A.~Ramirez, and S.~Speziali, ``{New AdS$_{2}$ backgrounds
  and $ \mathcal{N} $ = 4 conformal quantum mechanics},''
  \href{http://dx.doi.org/10.1007/JHEP03(2021)277}{{\em JHEP} {\bf 03} (2021)
  277}, \href{http://arxiv.org/abs/2011.00005}{{\tt arXiv:2011.00005
  [hep-th]}}.

\bibitem{Lozano:2020sae}
Y.~Lozano, C.~Nunez, A.~Ramirez, and S.~Speziali, ``{AdS$_{2}$ duals to ADHM
  quivers with Wilson lines},''
  \href{http://dx.doi.org/10.1007/JHEP03(2021)145}{{\em JHEP} {\bf 03} (2021)
  145}, \href{http://arxiv.org/abs/2011.13932}{{\tt arXiv:2011.13932
  [hep-th]}}.

\bibitem{Lozano:2021rmk}
Y.~Lozano, C.~Nunez, and A.~Ramirez, ``{$\text{AdS}_2\times \text{S}^2\times
  \text{CY}_2$ solutions in Type IIB with 8 supersymmetries},''
  \href{http://dx.doi.org/10.1007/JHEP04(2021)110}{{\em JHEP} {\bf 04} (2021)
  110}, \href{http://arxiv.org/abs/2101.04682}{{\tt arXiv:2101.04682
  [hep-th]}}.

\bibitem{Gubser:2002tv}
S.~S. Gubser, I.~R. Klebanov, and A.~M. Polyakov, ``{A Semiclassical limit of
  the gauge / string correspondence},''
  \href{http://dx.doi.org/10.1016/S0550-3213(02)00373-5}{{\em Nucl. Phys. B}
  {\bf 636} (2002)  99--114}, \href{http://arxiv.org/abs/hep-th/0204051}{{\tt
  arXiv:hep-th/0204051}}.

\bibitem{Kazakov:2004nh}
V.~A. Kazakov and K.~Zarembo, ``{Classical / quantum integrability in
  non-compact sector of AdS/CFT},''
  \href{http://dx.doi.org/10.1088/1126-6708/2004/10/060}{{\em JHEP} {\bf 10}
  (2004)  060}, \href{http://arxiv.org/abs/hep-th/0410105}{{\tt
  arXiv:hep-th/0410105}}.

\bibitem{Beisert:2003jj}
N.~Beisert, ``{The complete one loop dilatation operator of N=4 superYang-Mills
  theory},'' \href{http://dx.doi.org/10.1016/j.nuclphysb.2003.10.019}{{\em
  Nucl. Phys. B} {\bf 676} (2004)  3--42},
  \href{http://arxiv.org/abs/hep-th/0307015}{{\tt arXiv:hep-th/0307015}}.

\bibitem{Berenstein:2002jq}
D.~E. Berenstein, J.~M. Maldacena, and H.~S. Nastase, ``{Strings in flat space
  and pp waves from N=4 superYang-Mills},''
  \href{http://dx.doi.org/10.1088/1126-6708/2002/04/013}{{\em JHEP} {\bf 04}
  (2002)  013}, \href{http://arxiv.org/abs/hep-th/0202021}{{\tt
  arXiv:hep-th/0202021}}.

\end{thebibliography}\endgroup

\end{document}